\documentclass[graybox]{svmult}

\usepackage{mathptmx}       % selects Times Roman as basic font
\usepackage{helvet}         % selects Helvetica as sans-serif font
\usepackage{courier}        % selects Courier as typewriter font
\usepackage{type1cm}        % activate if the above 3 fonts are
\usepackage{makeidx}         % allows index generation
\usepackage{graphicx}        % standard LaTeX graphics tool
\usepackage{multicol}        % used for the two-column index
\usepackage[bottom]{footmisc}% places footnotes at page bottom
\usepackage{amssymb}

\usepackage{bm}% bold math
\usepackage{amsfonts}
\usepackage{latexsym}
\usepackage{amsmath}
\usepackage{psfrag}
\makeindex             % used for the subject index
\newcommand{\lb}{\label}

\newcommand{\bw}{\begin{widetext}}
\newcommand{\ew}{\end{widetext}}
\newcommand{\dv}{{\dot v}}
\newcommand{\da}{{\dot a}}

\newcommand{\ep}{{\varepsilon}}
\newcommand{\ddv}{{\ddot v}}

\newcommand{\bn}{{\rm bound}}
\newcommand{\ffi}{{\varphi}}

\newcommand{\be}{\begin{equation}}
\newcommand{\ee}{\end{equation}}
\newcommand{\ba}{\begin{align}}
\newcommand{\ea}{\end{align}}
\newcommand{\bea}{\begin{eqnarray}}
\newcommand{\eea}{\end{eqnarray}}
\begin{document}

\title*{Radiation reaction and energy-momentum conservation}
\author{Dmitri Gal'tsov}
\institute{Dmitri Gal'tsov\at Department of Physics, Moscow State
University, Russia; \email{galtsov@phys.msu.ru}  }
\maketitle

\abstract*{Each chapter should be preceded by an abstract (10--15 lines long)
that summarizes the content. The abstract will appear \textit{online} at
\url{www.SpringerLink.com} and be available with unrestricted access.
This allows unregistered users to read the abstract as a teaser for the
complete chapter. As a general rule the abstracts will not appear in
the printed version of your book unless it is the style of your particular
book or that of the series to which your book belongs.
Please use the 'starred' version of the new Springer \texttt{abstract}
command for typesetting the text of the online abstracts (cf. source file o
f this chapter template \texttt{abstract}) and include them with the source
files of your manuscript. Use the plain \texttt{abstract} command if the
abstract is also to appear in the printed version of the book.}

\abstract{We discuss subtle points of the momentum balance for
radiating particles in flat and curved space-time. An instantaneous
balance is obscured by the presence of the Schott term which is a
finite part of the bound field momentum. To establish the balance
one has to take into account the initial and final conditions for
acceleration, or to apply averaging. In curved space-time an
additional contribution arises from the tidal deformation of the
bound field. This force is shown to be the finite remnant from the
mass renormalization and it is different both form the radiation
recoil force and the Schott force. For radiation of
non-gravitational nature from point particles in curved space-time
the reaction force can be computed substituting the retarded field
directly to the equations of motion. Similar procedure is applicable
to gravitational radiation in vacuum space-time, but fails in the
non-vacuum case. The existence of the gravitational quasilocal
reaction force in this general case seems implausible, though it
still exists in the non-relativistic approximation. We also explain
the putative antidamping effect for gravitational radiation under
non-geodesic motion and derive the non-relativistic gravitational
quadrupole Schott term. Radiation reaction in curved space of
dimension other than four is also discussed}

\section{Introduction}
\label{sec:1} One of the major tasks of gravitational wave astronomy
is the precise theoretical prediction and observational measurement
of gravitational waveforms  from an inspiral fall of compact bodies
into the  supermassive black hole. This requires knowledge of the
orbits with account for gravitational radiation reaction. Radiation
gives rise to the reaction force which can be incorporated into the
equations of motion. The standard strategy to get the reaction force
consists in substitution of the  retarded field produced by the body
into its equation of motion. The resulting equation is believed to
give a correct description of the instantaneous effect of radiation
on the motion of the body. If the effect of reaction is small with
respect to the main force, one can treat it adiabatically. But when
the reaction force is not small, the situation is more subtle: the
balance equations involve not only the kinetic energy-momentum of
the body and that of radiation, but also a variable contribution of
the bound field. Generically, an instantaneous loss of the
energy-momentum by the body  is not equal to the energy-momentum
carried away by radiation.

This situation looks particularly simple in the flat space
Maxwell-Lorentz electrodynamics, where due to linearity of the
equations one can use the notion of the point-like particle
described by the delta-function. The Lorentz-Dirac equation obtained
by substituting the retarded field into the equation of motion and
performing the mass renormalization  is expected to describe the
motion of the particle with account for radiation loss. However, it
turns out that the momentum loss due to radiation gives only a part
of the reaction force. Although the initial system of the charges
and the Maxwell field obey the overall momentum conservation
equations, the Lorentz-Dirac equation, treated as the particle
equation of motion, violates the naively expected balance between
the particle momentum and the momentum of radiation. The difference
is given by the so-called Schott term, which is the third order
total derivative of the coordinate. A careful comparison of the
Dirac derivation \cite{Di38,IvSo48} and the Rohrlich analysis
\cite{FuRo60,Ro61,Ro65} has led Teitelboim \cite{Te70} (see also
\cite{Te80}) to interpret the Schott term as the finite contribution
of the bound (non-radiated) momentum of the charge remaining after
the mass renormalization (an explicit proof of this claim was given
in \cite{galpav}). This may look strange, since we are used to think
about the bound field as a stable Coulomb coat which is spherically
symmetric in the instantaneous rest frame or pancake shaped for a
relativistic charge. But this simple picture is valid only for
constant velocity. Once acceleration is non-zero, the
energy-momentum carried by the Coulomb coat becomes variable. The
most surprising fact is that this momentum is simply proportional to
the acceleration, and thus its derivative (the force) is the third
derivative of the particle coordinate. Of course, the split of the
total field into radiation and the bound field has to be done at any
distance from the charge, not only in the wave zone. This was given
by Rohrlich \cite{Ro65}, for more recent discussion see the book by
B. Kosyakov \cite{Ko07}.  It is worth noting that in this case  the
finite part of the reaction force is entirely given by the
time-antisymmetric part of the particle field (half-difference of
the retarded and advanced potentials).

Therefore the energy-momentum balance of the system consisting of
the accelerated charge and its Maxwell field includes three, but not
just two, ingredients: the particle momentum, the  momentum carried
by radiation, and the bound electromagnetic momentum. The radiation
momentum can be extracted both from the particle momentum and
indirectly from the bound momentum. This explains the origin of
radiation of the uniformly accelerated charge, in which case the
total reaction force is zero and thus the kinetic particle momentum
is constant. While the charge is undergoing a constant acceleration,
its bound electromagnetic momentum decreases and is transferred to
radiation. Physically, however, the acceleration has to start at
some moment and to finish at some moment, and during the stages of
acquiring and loosing the acceleration the bound momentum is
exchanged with the kinetic momentum. Therefore, the total
energy-momentum loss of the charge will be equal to the momentum
carried away by radiation. But an instantaneous balance is obscured
by the presence of the Schott term. Another simple situation is
periodic motion. Since the ambiguous Schott term is total
derivative, its contribution vanishes if one integrates over the
period, or, equivalently, averages over the period.

For radiation of linear fields of non-gravitational nature or of the
linearized gravitational field in curved space-time the situation is
more complicated   because there are no local conservation laws
unless the Killing symmetries are present, and because the tail term
can not be found in the closed form \cite{Dewi,hobbs} (for a review
see \cite{Po03}. Another new feature in this case is that the
reaction force contains a finite time-symmetric contribution coming
form the half-sum of the retarded and advanced potentials -- the
force found by DeWitt-DeWitt \cite{DWDW} in linearized gravity and
later rediscovered in the full General Relativity by Smith and Will
\cite{SmWi} (the WWSM force). This force, however, has nothing to do
with radiation and so the work done by this force is not expected to
contribute into the energy balance  for radiation. In the stationary
space-time the total energy of the particle and the field is
conserved, so one can expect that the energy balance between
radiation and the particle energy loss will hold integrally or in
average. In the case of axial symmetry, similar considerations apply
to an associated angular momentum. An explicit proof of this balance
for scalar, electromagnetic and linearized gravitational radiation
in the Kerr space-time was given in \cite{GaKer}. Apparently this
work  was not properly understood and was criticized in a number of
papers until 2005, when analogous calculations were performed by
other people (for a review and further references see \cite{Ta05}).

The case of linearized gravity is not entirely similar to other
linear fields in the curved background, however. In fact, linearized
gravity on the {\it non-vacuum} background is not a consistent
linear theory since the full Bianchi identities do  not allow for
the harmonic gauge necessary to locally disentangle the linearized
Einstein equations \cite{gaspist}. Physically this means that the
proper source of the gravitational radiation from the point particle
is not just its energy-momentum tensor, but it also includes
contribution from the perturbed source of the background. Thus the
total source is non-local, which raises doubts that there might
exist a local equation describing radiation reaction (here we mean
the non-locality stronger than that of the tail term, which is still
localized on the world-line of the point particle).

\section{Energy-momentum balance equation}
Consider the  interacting system of N point charges and the Maxwell
field in Minkowski space-time which is described by the coupled
system of equations
 \be
 \partial_\nu F^{\mu\nu}=4\pi\int \sum_{a=1}^N {\dot z}^\mu
\delta(x-z_a(\tau))d\tau,\qquad
   m^0_a \ddot z^\mu_a =e_a F^{\mu}_{\;\;\nu}{\dot z}_a^\nu.
  \ee
 It has $3N+\infty$ degrees of freedom, where $\infty$ stands for
 the Maxwell field.
  The corresponding energy-momentum conservation equation is
 \be\partial_\nu ( \stackrel{m}{T}{\!\!}^{\mu\nu}+
\stackrel{F}{T}{\!\!}^{\mu\nu})=0,\lb{totem}
 \ee
  where
 \be \stackrel{m}{T}{\!\!}^{\mu\nu}=\int\sum_{a=1}^N m^0_a{\dot z}^\mu
{\dot z}^\nu \delta(x-z_a(\tau))d\tau,\quad
 %\ee
%is the particles stress-tensor and
 % \be
 \stackrel{F}{T}{\!\!}^{\mu\nu}
=\frac1{4\pi}\left(F^{\mu\lambda} F_{\lambda}^{\;\nu}
+\frac{\eta^{\mu\nu}}4 F^{\alpha\beta}F_{\alpha\beta}\right).  \ee
We have introduced the bare masses $m^0_a$ in anticipation of
mass-renormalization. Since the Maxwell equation is linear, one can
decompose the total field in the vicinity of any given charge $e_a$
into the sum of the field generated by the
 other $N-1$ charges, $F_{\rm ext}^{\mu\nu}$ (regular at its
 location) and the retarded field of $e_a,\;F_{a\,{\rm ret}}^{\mu\nu}$.
In spite of the fact that the total field acting on $e_a,\;
F_a^{\mu\nu}=F_{\rm ext}^{\mu\nu}+F_{a\,{\rm ret}}^{\mu\nu}$
diverges at $x^\mu=z_a^\mu$, energy-momentum conservation is ensured
by the equations of motion. The required mass renormalization does
not change this statement.

    The field $F_{a\,{\rm ret}}^{\mu\nu}$ describes radiation
and the bound field, which both contribute to the field
energy-momentum. The overall conservation equation does not
distinguish between these two parts, so an additional analysis is
needed. The retarded potential at a given point $x$ of space-time
depends on the world-line variables taken at the moment of proper
time $s_{\rm ret}(x)$ defined as the solution to the equation
 \be R^{\mu} R_{\mu}=0,\quad
R^{\mu}=x^{\mu}-z^{\mu}(s_{\rm ret}),
 \ee
 satisfying $x^0>z^0$. The
advanced  solution to the same equation with $z^0>x^0$ refers to the
advanced proper time $s_{\rm adv}(x)$. Introducing the invariant
distance
 \be\label{ro}\rho=v_{\mu}(s_{\rm ret})R^{\mu},\quad
v^\mu=\frac{dz^\mu}{ds},
 \ee which is equal to the spatial distance
$|\mathbf{R}|=|\mathbf{x}-\mathbf{z}(s_{\rm ret})|$ between the
points of emission and  observation  in the momentarily co-moving
Lorenz frame at the time moment $x^0=z^0(s_{\rm ret})$, one can
present the retarded potential as  (we omit the index $a$):
 \be
    A^{\mu}_{\rm ret}(x)=\frac{e v^{\mu}}{\rho}\Big|_{s_{\rm
    ret}(x)}.
    \ee
Introduce the normalized null vector $c^{\mu}=R^{\mu}/\rho$, such
that $vc=1$, and the  unit space-like vector
$u^\mu=c^{\mu}-v^{\mu},\; u^2=-1$. The following differentiation
rules then hold: \be \lb{ccc} c_{\mu}=\partial_{\mu} s_{\rm
    ret}(x), \quad \partial_{\mu} \rho =v_{\mu}+ \lambda c_{\mu},
    \quad \partial_{\mu} c^{\nu}=\frac{1}{\rho}\left(\delta _{\mu}^{\nu}-
v_{\mu}c^{\nu}-c_{\mu}v^{\nu}-\lambda c_{\mu}c^{\nu},
 \right)
 \ee
where $\lambda=\dot{\rho}=\rho (ac)-1.$ The retarded field strength
will read: \be
 F_{\rm ret}^{\mu
 \nu}=\frac{e \left(\rho(ac)-1\right)}{\rho^2}v^{[\mu}c_{\nu]}-
 \frac{e }{\rho}a^{[\mu}c_{\nu]}.
\ee
  The retarded potential in Minkowski space admits a
natural decomposition with respect to T-parity: \be \lb{aret}
A^\mu_{\rm ret}=A^\mu_{\rm self}+A^\mu_{\rm rad},\ee where the
radiative part $ A^\mu_{{\rm rad}}=\frac12\left(A^\mu_{{\rm
ret}}-A^\mu_{{\rm adv}}\right)$ obeys an homogeneous wave equation,
while the self part $ A^\mu_{{\rm self}}=\frac12\left(A^\mu_{{\rm
ret}}+A^\mu_{{\rm adv}}\right)$ has a source at $\  x=z(s)$.  One
could expect that only T-symmetric $ A^\mu_{{\rm self}}$ corresponds
to the {\it  bound} field, but it is not so. For an accelerated
charge  the situation is more subtle.
 \subsection{Decomposition of the stress tensor}
Constructing   the energy-momentum tensor
$\stackrel{F}{T}{\!\!}^{\mu\nu}$ with the retarded field $F_{\rm
ret}^{\mu\nu}$, one finds that it admits a natural decomposition:
\be
\stackrel{F}{T}{\!\!}^{\mu\nu}=\stackrel{F}{T}{\!\!}^{\mu\nu}_{\rm
emit}+\stackrel{F}{T}{\!\!}^{\mu\nu}_{\rm bound},\ee where the first
term is selected by its dependence on $\rho$ as $\rho^{-2}$:
\begin{equation}\label{SET4Dradiat}
\stackrel{F}{T}{\!\!}^{\mu\nu}_{\rm
emit}=-\frac{((ac)^2+a^2)c_{\mu}c_{\nu}}{\rho^2},
\end{equation}
while the second contains higher powers of $\rho^{-1}$:\be
\label{Tbound4c}
 \stackrel{F}{T}{\!\!}^{\mu\nu}_{\rm
 bound} = \frac{a^{(\mu}c^{\nu)}+2(ac)c^{\mu}c^{\nu}-
 (ac)v^{(\mu}c^{\nu)}}{\rho^3} \nonumber +
 \frac{v^{(\mu}c^{\nu)}-c^{\mu}c^{\nu}- \eta^{\mu
 \nu}}{\rho^4}/2,
\ee  where  symmetrization without $1/2$ is understood.

The ``emit'' part (\ref{SET4Dradiat}) has the following properties:
\begin{itemize}\item it is the tensor product
of two null vectors $c^\mu$, \item it is traceless, \item it falls
down as $|{\mathbf x}|^{-2}$ when $|{\mathbf x}|\to \infty$,
\item as follows from the differentiation rules (\ref{ccc}),
it is divergence-free without assuming the validity of the equations
of motion:
\begin{equation}\label{radconserv}
\partial_{\mu}\stackrel{F}{T}{\!\!}^{\mu\nu}_{\rm emit}=0.
\end{equation}
\end{itemize}
All these features indicate that $T^{\mu\nu}_{\rm emit}$ describes
the outgoing radiation.

Since the total energy-momentum tensor including the contribution of
charges is (on shell) divergence free, with account for
(\ref{radconserv}) we find that
\begin{equation}\label{boundconserv}
\partial_{\mu}\stackrel{F}{T}{\!\!}^{\mu\nu}_{\rm bound}+
\partial_{\mu}\stackrel{m}{T}{\!\!}^{\mu\nu}=0,
\end{equation}
so  the bound field momentum can be exchanged with the particle
momentum. Note, that outside the world-line the bound stress tensor
is divergence-free. It is also worth noting, that Eq.
(\ref{radconserv}) does not mean that there is no reaction force
acting on a particle which counterbalances the emitted momentum.

Consider now the total balance of forces. The conservation of the
total four-momentum (\ref{totem}) implies that the sum of the
mechanical momentum and the momentum carried by the electromagnetic
field is constant (for simplicity we do not include the external
field): \be \frac{d p^\mu_{\rm mech}}{ds}+\frac{d p^\mu_{\rm
em}}{ds}=0. \ee Here the mechanical part is proportional to the bare
mass of the charge \be p^\mu_{\rm
mech}=\int\stackrel{m}{T}{\!\!}^{\mu\nu}d\Sigma_\nu=m^0 v^\mu,\ee
while the field part is given by \be p^\mu_{\rm em} =\int
\stackrel{F}{T}{\!\!}^{\mu\nu}d\Sigma_\nu,\ee where integration of
the electromagnetic stress tensor is performed over a space-like
hypersurface whose choice will be specified later on. It has to be
emphasized that  the stress tensor of the electromagnetic field is
constructed in terms of the physical retarded field.  According to
the above splitting, we can write \bea \frac{d
p^\mu_{\rm mech}}{ds}&=&f^\mu_{\rm emit}+f^\mu_{\rm bound},\\
f^\mu_{\rm emit}=-\int \stackrel{F}{T}{\!\!}^{\mu\nu}_{\rm emit}
d\Sigma_\nu, & \quad & f^\mu_{\rm bound}=-\int
\stackrel{F}{T}{\!\!}^{\mu\nu}_{\rm bound}d\Sigma_\nu.\eea

On the other hand, the derivative of the bare mechanical momentum of
the charge can be found substituting the retarded field in the
equation of motion. In this case it is useful to decompose the
retarded field according to (\ref{aret}), obtaining another split of
the mechanical momentum: \be \frac{d p^\mu_{\rm mech}}{ds} = e
F^{\mu\nu}_{\rm ret}v_\nu = e\left( F^{\mu\nu}_{\rm
self}+F^{\mu\nu}_{\rm rad}\right)v_\nu  = f^\mu_{\rm
self}+f^\mu_{\rm rad}. \ee

Now, somewhat unexpectedly, $ f^\mu_{\rm rad}\neq f^\mu_{\rm emit}$
and $ f^\mu_{\rm self}\neq f^\mu_{\rm bound}$, the difference being
called  the Schott term \cite{galpav}:\be f^\mu_{\rm rad} =
f^\mu_{\rm emit}+ f^\mu_{\rm Schott}, \quad f^\mu_{\rm self} =
f^\mu_{\rm bound}- f^\mu_{\rm Schott}.\ee Clearly, \be
\lb{consistency}f^\mu_{\rm self}+f^\mu_{\rm rad}=f^\mu_{\rm
bound}+f^\mu_{\rm emit},\ee as expected.  Note that both $f^\mu_{\rm
bound}$ and $f^\mu_{\rm self}$ contain divergences which mutually
cancel  in  Eq. (\ref{consistency}).
%\section{World-line calculation: point splitting}

The forces $f^\mu_{\rm self}$ and $f^\mu_{\rm rad}$ can be found
using the  Green functions \cite{IvSo48} \be \lb{Gsr} G_{\rm
self}(Z)=\delta{( Z^2)},\quad G_{\rm
rad}(Z)=\frac{Z^0}{|Z^0|}\delta{( Z^2)}, \ee where
$Z^\mu=Z^\mu(s,s')=z^\mu(s)-z'^\mu(s')$. Substituting the value of
the electromagnetic field generated by the charge on its world-line
one obtains   \be f^\mu(s)=2
e^2\int\;Z^{[\mu}(s,s')v^{\nu]}(s')v_\nu(s)\frac{d}{d Z^2} G(Z) ds',
\ee  for both $f^\mu_{\rm self}$ and $f^\mu_{\rm rad}$. Due to
delta-functions,  only a finite number of Taylor expansion terms in
$\sigma=s-s'$ contribute to the integral. In the four-dimensional
case it is sufficient to retain the terms up to $\sigma^3$:
\be\lb{exp} 2Z^{[\mu}(s,s')v^{\nu]}(s')v_\nu(s)=\dv^\mu\sigma^2-
\frac23(\ddv^\mu+v^\mu\dv^2 )\sigma^3 +O(\sigma^4). \ee Taking into
account that $Z^2=\sigma^2+O(\sigma^4)$, the leading terms in the
expansions of derivatives of the Green functions will be \be
\lb{sings} \frac{d}{d Z^2} G_{\rm self}(Z) = \frac{d}{d
\sigma^2}\delta(\sigma^2), \quad \frac{d}{d Z^2} G_{\rm rad}(Z) =
\frac{d}{d
\sigma^2}\left(\frac{\sigma}{|\sigma|}\delta(\sigma^2)\right). \ee
Regularizing the delta functions of $\sigma^2 $ by point-splitting
\be \delta(\sigma^2)=\lim_{\ep\to +0}
\delta(\sigma^2-\ep^2)=\lim_{\ep\to
+0}\frac{\delta(\sigma-\ep)+\delta(\sigma+\ep)}{2\ep},\ee with a
prescription that the limit should be taken  after evaluating
 the integrals, one finds  \be  \lb{fself} f^\mu_{\rm
self} = -  \frac{e^2 }{2 \ep} a^\mu,\quad f^\mu_{\rm rad} =
\frac{2e^2}3(v^\mu a^2+\da^\mu).\ee After the mass renormalization,
$ m_0+\frac1{2 \ep }=m $ we get the Lorentz-Dirac equation \be m
a^\mu= \frac{2e^2}3(v^\mu a^2+\da^\mu). \ee The first terms at the
right hand side  is equal to the derivative of the momentum carried
away be radiation, \be f^{\mu}_{\rm emit}=-\frac{dp^\mu_{\rm
emit}}{ds}=\frac{2e^2}{3} a^2 v^{\mu}.\ee Its independent evaluation
by integration of $ \stackrel{F}{T}{\!\!}^{\mu\nu}_{\rm emit}$ can
be found in \cite{galpav}. The second total derivative term is the
Schott term. It is worth noting, that within the local calculation,
the Schott term originates form the T-odd part of the retarded
field.
\subsection{Bound momentum}
An explicit evaluation of the bound momentum  associated with a
given moment of proper time $s$ on the particle world-line, \be
\label{momentdefinition} p_{\rm bound}^{\mu}(s)=\int
\limits_{\Sigma(s)}\stackrel{F}{T}{\!\!}^{\mu\nu}_{\rm
bound}d\Sigma_{\nu}  \ee was given in \cite{galpav}, which we follow
here. First of all one has to choose the space-like hypersurface
$\Sigma(s)$ intersecting the world line at $x^\mu=z^\mu(s)$. A
convenient choice will be the hyperplane orthogonal to the
world-line \be\label{hypersigma}
v_\mu(s)\left(x^\mu-z^\mu(s)\right)=0. \ee
\begin{figure}
\centerline{\psfrag{s1}{\LARGE $ s_{1}$}    \psfrag{s2}{\LARGE $
s_{2}$}   \psfrag{Sie}{\LARGE $S_{\varepsilon}$}
\psfrag{Sis1}{\LARGE $ \Sigma(s_1)$}    \psfrag{dYRs1}{\LARGE $
\partial Y_R(s_1)$}
\psfrag{Yis2}{\LARGE $ Y(s_2)$}\psfrag{SiR}{\LARGE $ S_R$}
\psfrag{zmis}{\LARGE $ z^{\mu}(s)$}  \psfrag{dYes1}{\LARGE $\partial
Y_{\varepsilon}(s_1)$}
\includegraphics
[angle=270,width=15cm]{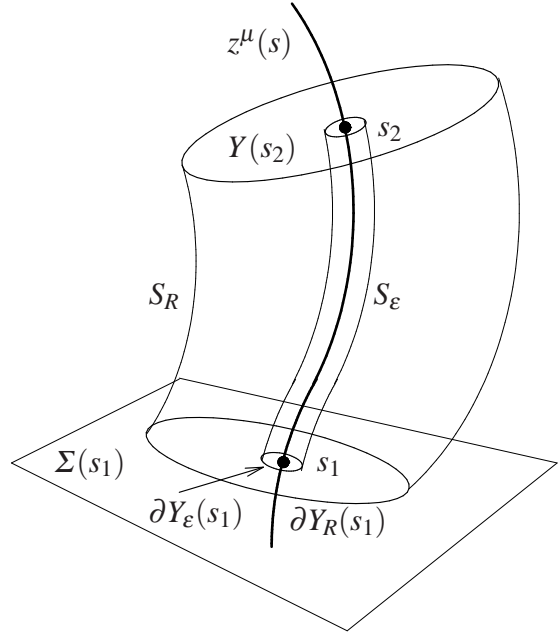}}\caption{Integration  of the bound
electromagnetic  momentum. Here  $\Sigma(s_1)$ is the space-like
hyperplane transverse to the world-line  $z^\mu(s)$ intersecting it
at the proper time $s_1$ (similarly $\Sigma(s_2)$). The
hypersurfaces $S_{\varepsilon}$ and $S_R$ are small and large tubes
around the world-line formed by sequences of the 2-spheres $\partial
Y_{\varepsilon}(s)$ and $\partial Y_R(s)$ for $s\in [s_1, \,s_2]$.
The domain $Y(s_2)\subset \Sigma(s_2)$ (similarly $Y(s_1)$) is the
3-annulus between $\partial Y_R(s_2)$ and $\partial
Y_{\varepsilon}(s_2)$.} \label{figur2}
\end{figure}
The integral (\ref{momentdefinition}) is divergent on the world
line. We introduce the small length parameter $\ep$, the radius of
the 2-sphere $\partial Y_\ep(s)$ (Fig. 1), defined as the
intersection of the hyperplane (\ref{hypersigma}) with the
hyperboloid $ (x-z(s))^2=-\varepsilon^2$. We also introduce the
sphere $\partial Y_R(s)$ of the large radius $R$ defined as the
intersection of $\Sigma(s)$ with the hyperboloid $(x-z(s))^2=- R^2$.
The total field momentum can be obtained as the limit $\ep\to 0,\;
R\to \infty$ of the integral over the domain $Y(s)\subset \Sigma(s)$
between the boundaries $\partial Y_\ep(s)$ and $\partial Y_R(s)$.

Let us evaluate the variation of this quantity between the moments
$s_1$ and $s_2$ of the proper time on  the world-line of the charge
\be \Delta p^\mu_{\rm em}=\int \limits_{Y(s_2)}T^{\mu
\nu}d\Sigma_{\nu}-\int \limits_{Y(s_1)}T^{\mu \nu}d\Sigma_{\nu}.\ee
For the bound momentum it is convenient to consider the tubes
$S_\ep$ and $S_R$ formed as sequences of the spheres $\partial
Y_\ep(s)$ and $\partial Y_R(s)$ on the interval $s\in [s_1,\,s_2]$
and to transform this quantity to \be \Delta p^\mu_\bn=\int
\limits_{S_R }T^{\mu \nu}_\bn dS_{\nu}-\int \limits_{S_\ep}T^{\mu
\nu}_\bn dS_{\nu}\ee in view of the conservation equation for
$T^{\mu\nu}_\bn $ outside the world-line (see the remark after Eq.
(\ref{boundconserv})). Here the integration elements $dS_\nu$ are
directed outwards  to the world-line. The contribution from the
distant surface $S_R$ vanishes if one assumes that the  acceleration
is zero in the limit $s\to-\infty$~\cite{Te70}. This  is
non-trivial: though the stress tensor (\ref{Tbound4c}) decays as
$R^{-3}$ at spatial infinity, the corresponding flux does not vanish
{\em a priori}, because the surface element contains a term
(proportional to the acceleration) which asymptotically grows as
$R^3$. As  a consequence, the surviving term will be proportional to
the acceleration taken at the moment $s_{\rm ret}$ of the proper
time, where $s_{\rm ret}\to -\infty$  in the limit $R\to\infty$.
Finally we are left with the integral over the inner boundary
$S_\ep$ only. To find an integration measure on $S_\ep$ we foliate
the space-time domain shown in Fig. 1 by the hypersurfaces
$\Sigma(s)$ parameterized by the spherical coordinates $r,
\theta_1=\theta, \theta_2=\ffi$. Introducing the unit space-like
vector $n^\mu (s,\theta_i),\,n_\mu n^\mu=-1$  transverse to $v^\mu$,
we use the coordinate transformation  $x^{\mu}=z^{\mu}(s)+r
n^{\mu}(s,\theta_{i} )$. The induced metric on $S_\ep$ reads
$dS_{\mu}=\varepsilon^{2}[1-\varepsilon(an)]n_{\mu} ds d\Omega$, and
hence  \be\lb{dps}\frac{dp^\mu_{\rm bound}}{ds}=-\int_{S_\ep} \ep^2
[1-\ep(an)]\stackrel{F}{T}{\!\!}^{\mu\nu}_{\rm bound}n_\nu
d\Omega,\ee  where the limit $\ep\to 0$ has to be taken. One has to
expand $\stackrel{F}{T}{\!\!}^{\mu\nu}_{\rm bound}$ in terms of
$\ep$. In fact, the energy-momentum tensor depends on the space-time
point $x^\mu$ through the quantity $\rho$, depending directly on
$x^\mu$, and also through the retarded proper time $s_{\rm ret}$. We
have to express the resulting quantity as a function of the proper
time $s$ corresponding to the intersection of the world-line with
the space-like hypersurface.  We write $T_{\mu \nu}^{\rm bound}$ in
terms of  the null vector $R^\mu=c^\mu\rho$:
\begin{align}\label{Tbound4R}
\frac{4\pi}{e^{2}} T^{\mu \nu}_{\rm bound} = \frac
 {a^{(\mu}R^{\nu)}}{\rho^4}+\frac{\left(2(aR)-1\right)R^{\mu}R^{\nu}}{\rho^6}+
  \frac{\left(1-(aR)\right)v^{(\mu}R^{\nu)}}{\rho^5}
  -\frac{\eta^{\mu \nu}}{2\rho^4},
\end{align}
and expand $R^\mu$ as
\begin{align}\label{R(sig)}
R^\mu=x^\mu-z^\mu(s_{\rm{ret}})= \varepsilon n^\mu+v^\mu
 \sigma-\frac{1}{2}a^\mu \sigma^2+\frac{1}{6}\dot{a}^\mu \sigma^3
 +\mathcal{O}(\sigma^4),
\end{align}
where  $\sigma=s-s_{\rm{ret}}>0$ and all the vectors  are taken at
$s$. This   expansion in powers of $\sigma$ has to be rewritten in
terms of $\varepsilon$. The relation between the two can be found
from the condition $R^2=0$:
\begin{align}\label{sigma(eps)}
\sigma=\varepsilon+\frac{an}{2}\varepsilon^2+\left( 9(an)^2+
 a^2- 4\dot{a}n\right)\frac{\ep^3}{24}+\mathcal{O}(\ep^4).
\end{align}
Substituting this into  Eq.~(\ref{R(sig)}) and further to
(\ref{Tbound4R}) we find:
   \begin{align}\label{N4(eps)} &\Delta
p^\mu_\bn=\frac{e^2}{4\pi}\int\limits_{s_1}^{s_2} ds
  \left\{\frac{-n^{\mu}}{2\varepsilon^2}+\frac{a^{\mu}}{2
\varepsilon}+\left[\left((an)^2+a^2/3\right)
v^{\mu}+\right.\right.\nonumber\\&+\left.\left.\left((an)^2+a^2/2\right)
n^{\mu}- 2\dot{a}^{\mu}/3  + 3(an)a^{\mu}/4\right]\right\}
 d\Omega.\end{align} The leading divergent term $1/\ep^2$ disappears after
angular integration. Thus we obtain \cite{galpav}:
\begin{align}\label{boundforce}
f^{\mu}_\bn
=-\frac{dp^\mu_\bn}{ds}=-\frac{e^2a^\mu}{2\ep}+\frac{2e^2}{3}
\da^{\mu}.
\end{align}
Here the first divergent term has to be absorbed by the
renormalization of mass, while the second is the finite Schott term.
Comparing this with (\ref{fself}) we confirm  the identity
(\ref{consistency}).   Note that {\em a priori} the regularization
parameter  $\ep$ here (the radius of the small tube) is not related
to the splitting parameter   of the delta-function in the previous
local force calculation. But actually they give the same form for
divergent terms, for which reason we use the same symbol $\ep$ for
both of them. With this convention, the divergent terms in the
momentum conservation identity (\ref{consistency}) mutually cancel.

The significance of the Schott term in the balance of momentum
between the radiating charge and the emitted radiation is not always
recognized in  the text-books on classical electrodynamics. Instead,
its presence is often interpreted as a drawback of classical theory,
since formally it may lead to self-accelerating solutions. Meanwhile
such  solutions must be discriminated as unphysical since they do
not satisfy proper initial/final conditions which should be imposed
on the third order equation of motion \cite{IvSo48}. From the above
analysis it is clear that the Lorentz-Dirac equation, although
formulated in terms of particle variables, actually describe the
composite system consisting of a charge and its bound
electromagnetic momentum.  The redefinition of the particle momentum
joining to it the bound electromagnetic momentum obscures the
problem of interpretation. It is better think of the Schott term as
the field degree of freedom and interpret  the Lorentz-Dirac
equation as the momentum balance equation for the total system
including the electromagnetic field.

An instantaneous momentum balance is not just the balance between
the particle and radiation, the energy-momentum can be also
transferred between the particle and the  field coat bound to it.
Or, the radiated momentum is not always taken from the mechanical
momentum of the charge, but by virtue of the Schott term, it can be
extracted from the field coat too. This is what happens in the case
of the uniformly accelerated charge, when the total reaction force
instantaneously is zero, while radiation carries the momentum away
at a constant rate. The balance in ensured by the Schott term.
However, the constant acceleration can not last infinitely long. One
has to consider the switching on/off processes in order to
understand that finally the energy-momentum of radiation is taken
from the particle. This consideration clarifies the necessity of the
time averaging or integration over time needed to establish the
momentum balance between radiation and the source particle. The
equation of motion including the reaction force instantaneously does
not imply the equality of the radiative momentum loss and the
particle momentum. This feature is general enough, it is also
applicable to radiation of non-gravitational nature from  particles
moving along the geodesics  in curved space-time, as well as to the
gravitational radiation.

\subsection{The rest frame (non-relativistic limit)}
In the rest frame of a charge the recoil force has no spatial
component. This is due to the fact that radiation in two opposite
directions is the same so that the spatial momentum is not lost by
radiation, though the energy is lost. Hence the total spatial
component of the reaction force is presented by the Schott term,
namely \be{\bf f_{\rm Schott}}=\frac23 e^2 {\bf \dot a}.\ee The work
done by this force \be\int{\bf f_{\rm Schott}}\cdot{\bf
v}dt=\int\frac23 e^2 {\bf \dot a}\cdot{\bf v}dt=-\int\frac23 e^2
{\bf a}^2 dt+ \;{\rm boundary\;\;\; terms}\ee correctly reproduces
the radiative loss in the rest frame. (Boundary terms should vanish
by appropriate asymptotic switching on/off or periodicity
conditions).
\section{Flat dimensions other than four}
Recent interest to models with large extra dimensions motivates the
study of radiation and radiation reaction in dimensions other than
four. It turns out that the radiation picture is substantially
different in even and odd dimensions because of the different
structure of the retarded Green's functions for massless fields in
the coordinate representation \cite{Ga01} (they still look similarly
in all dimensions in the momentum representation). In even
dimensions the retarded potential is localized on the past light
cone (Huygens principle) so the situation is qualitatively similar
to that in the 4D case. In odd dimensions it is non-zero also inside
the past null cone though radiation still propagates along the null
rays. In 3D, for instance, the scalar Green's function reads \be
G_{\rm ret}^{3D}(X)=\vartheta(X^0)\vartheta(X^2)(X^2)^{(-1/2)},\quad
X^\mu=x^\mu-x'^\mu.\ee It does not-contain the ``direct'' part
singular on the light cone. Green's functions in higher odd
dimensions $D=2n+1$ can be obtained by the recurrent relation
\cite{Ga01,GScurv} \be G_{\rm ret}^{2n+1}(X)\sim
\frac{dG^{2n-1}_{\rm ret}}{dX^2}\vspace{-.3cm}\ee In particular, in
5D \be G_{\rm ret}^{5D}(X)\sim
\vartheta(X^0)\left(\frac{\delta(X^2)}{(X^2)^{1/2}}-\frac{1}{2}
\frac{\vartheta(X^2)}{(X^2)^{3/2}}\right) \ee both the direct and
the tail parts are present.  It turns out that the direct part
regularizes the tail contribution to the field stress proportional
to the derivative of $G$ which otherwise would be singular outside
the world line.

 In even  dimensional space-times the split of the retarded
potentials into the time symmetric and the radiative  parts  leads
to purely divergent self-force and a finite radiative part: \be
 f^\mu_{\rm self}=f^\mu_{\rm div},\quad f^\mu_{\rm rad}={\rm
 finite}.\ee
Divergent terms are Lagrangian type and can be absorbed by
introducing suitable counterterms. Since the Coulomb dependence is
more singular at the location of the source in higher dimensions,
the self-action gives rise to larger number of divergent terms
$1/\ep^n$, where $n$ is changing from unity to the integer part of
$D/2-1$. The highest divergence is absorbed by the renormalization
of mass, while  to absorb other divergencies  the counterterms are
needed depending on higher derivative of the velocity. These are not
present in the initial action, so higher-dimensional classical
theories are not renormalizable. In 6D one has two divergent terms
(which in terms of the field split correspond to $f_{\rm self}$
\cite{Ko99}: \be f^\mu_{\rm div}= -\frac{1}{6 \ep^3}
a_{\mu}+\frac{1}{2\ep}\left(\frac{3}{4}
v_{\mu}(\dot{a}a)+\frac{3}{8} a^2
a_{\mu}+\frac{1}{4}\ddot{a}_{\mu}\right), \ee the leading being
eliminated by the mass renormalization and the subleading requiring
the counterterm \cite{Ko99}\be S_1=-\kappa^{(1)}_0 \int ( \ddot z
)^2 ds, \ee which leads to the Frenet-Serret dynamics \cite{Capo}
unless the renormalized value  $\kappa^{(1)}=0$. For each two
space-time dimensions one new higher-derivative counterterm is
needed  to absorb divergencies.

 The split of the field stress-tensor built with the retarded field  into
the sum of the emitted and bound terms is also possible in all even
dimensions, and one  always has the relation (\ref{consistency}). In
6D, e.g., the radiation recoil force in 6D is \be f^{\mu}_{\rm
emit}= \frac{4}{45}e^2\left( \dot{a}^2v^{\mu}+\frac{2}{21} (a
\dot{a})a^{\mu}- \frac{2}{9} a^4 v^{\mu}-\frac{2}{105} a^2
\dot{a}^{\mu}\right),\ee and the Schott terms is \be f^{\mu}_{\rm
Schott}=-\frac{4 e^2}{45}\left( \dddot{a}^{\mu}+\frac{16}{7}a^2
\dot{a}^{\mu}+\frac{60}{7}(a \dot{a})a_{\mu} + 4\dot{a}^2 v^{\mu}
+4(a\ddot{a})v^{\mu}\right),\ee the sum of two being orthogonal to
the 6-velocity. The Schott terms is again given by the finite part
of the integrated bound momentum.

In odd dimensions one always have tail terms. Split of the retarded
field into  the self-  and the  radiative parts is always possible,
and the substitution into the equations of motion leads to divergent
and finite terms. Divergent ones can again be absorbed introducing
the counterterms. The split of the stress-tensor into the emitted
and the bound parts is more subtle. The situation is obscured by the
fact that though the free field is massless and thus propagates
along the null cone, the full retarded potential fills the interior
of the past light cone. Still one is able to obtain the general
formula for radiation momentum which is no more associated with the
retarded proper time on the world-line \cite{GaSp_prep}.

\section{Local method for curved  space-time}
An approach initiated by DeWitt and Brehme and further applied to
linearized gravity in \cite{mino} appeals to computation of the
integral of the stress-tensor in the world-tube surrounding the
world-line. This is similar to our calculation of the bound
momentum. However, in curved space-time the split of the stress
tensor into the emitted and  bound parts becomes problematic,  so
the complete analysis of the balance between radiation, the kinetic
momentum and the bound momentum is not available. Meanwhile, to
compute the total reaction force one can use much a simpler
calculation substituting the retarded field directly into the
equations of motion \cite{gaspist}. This approach was also
formulated in higher even-dimensional space-time in the paper
\cite{GScurv} which we follow here.

\subsection{Hadamard expansion in any dimensions}\label{qq22sc}
As in $D=4$ \cite{Dewi}, the curved space Green's functions for
massless fields in other dimensions  can be constructed starting
with the Hadamard solution. For simplicity we consider here the
scalar case. The scalar Hadamard   Green's function $G_H(x,x')$ is a
solution of the homogeneous  wave equation $
 \Box_x G_H(x,x') =0,
$ where $\Box=g_{\mu\nu}\nabla^{\mu} \nabla^{\nu}$. The procedure
consists in expanding $G_H(x,x')$ in terms of the Synge world
function $ \sigma(x,x')$.
  For $D=4$ the Hadamard
expansion contains two terms singular in $\sigma$, namely,
$\sigma^{-1}$ and $\ln \sigma$. In higher dimensions one has to add
other singular terms, and by dimensionality it is easy to guess that
each dimension introduces   an additional factor $\sigma^{-1/2}$.
Thus, the Hadamard expansion in $D=2d$ dimensions ($d\geqslant 3/2$
is integer or half-integer) generically must read
\begin{equation}\label{phi3} G_H(x,x') =\frac{1}{(2\pi)^{d}}
\left[\sum_{n=1}^{D}g_n \sigma^{1-n/2}+v \ln \sigma\right],
\end{equation}
where $g_n=g_n(x,x'),\; v=v(x,x')$ are  two-point  functions. It can
be  shown, that in odd dimensions we actually have only odd powers
of $\sigma^{-1/2}$, and in even dimensions --- only even powers,
that is, an expansion in terms of inverse integer powers of
$\sigma$. The logarithmic term  is present only in even dimensions.

Substituting  (\ref{phi3}) into the wave equation, in the leading
singular order we will have: $g_{D}=\Delta^{1/2}.$ In the next to
leading order we obtain the equation:
\begin{equation} \label{phi8}
  2 \, \partial_\mu g_{D-1} \sigma^\mu +g_{ D-1 }\Box\sigma -  (D-1)
g_{ D-1 }=0.
\end{equation}
which does not have analytic solutions, so $g_{D-1}=0$.  For $D=3$
this means the absence of the logarithmic term.
  Similarly,
considering the equation for $g_{(D-1-2k)},\;k \in \mathbb{N}$  we
find
$
g_{D-1-2k}=0.
$
This means that for an even-dimensional space-time the Hadamard
Green's function contains only integer negative powers of
 $\sigma$ plus logarithm and a regular part, while in the odd-dimensional case -- only
half-integer powers of $\sigma$ plus a regular part.

 For the sequence of Green's
functions in the flat space-time, the one in  $D+2$ dimension is
proportional to the derivative  of the Green's function in the twice
preceding dimension $D$. In fact, in even dimensions the symmetric
Green's function is the derivative of the order $d-2$ of the
delta-function: $G^{D}\sim
\delta^{d-2}(\sigma),\quad\sigma=(x-x')^2/2 $ and thus, $G^{D+2}\sim
d G^{D}/d\sigma$. Applying  regularization,
$\delta((x-x')^2)=\lim_{\varepsilon \to +0}\delta(
|(x-x')^2|-\varepsilon^2),$ we obtain
\begin{equation}G^{D+2}
 \sim d G^{D}/d\varepsilon^2.\end{equation}
 This relation has a
consequence that the Laurent expansion of the Lorentz-Dirac force in
terms of $\varepsilon$ in the even-dimensional Minkowski space has
only odd negative powers, and no even terms. So the number of
divergent terms in the self-action increases by one for each next
even dimension. In curved space passing to the standard notation
notation for $g_n$  we have
\begin{equation} \label{red3}
G_H=\frac{1}{(2\pi)^{d}}\left(\sum_{k=0}^{d-2}\frac{u_k}{\sigma^{d-1-k}}+
v \ln\sigma + w\right),
\end{equation}
where $u_0 =\Delta^{1/2}$ and we denoted $v= u_{d-1},\;w= u_{d}$.
Applying  $\Box$  with respect to $x^{\mu}$, we obtain  the  system
of recurrent differential equations for $u_i (x,x')$ . Integrating
them along the geodesic connecting the points $x,x'$, one can
uniquely express $u_1(x,x')$ through $u_0(x,x')$. Furthermore, $u_2$
is expressed through $u_1$, etc.

To relate the coefficient functions $u^{D}_i(x,x')$ in different
dimensions we first observe that $u_0= \Delta^{1/2}$ for any $D$. It
is worth noting, however, that in the expansion in terms of $\sigma$
\begin{equation} \label{vVleckexp}
u_0^{D}=\Delta^{1/2}=1+1/12\,R_{\alpha\beta}\sigma^{\alpha}\sigma^{\beta}+...
\end{equation}
where the tensor indices $\alpha,\,\beta$  run the values
corresponding to $D$.   With this in mind, we can write $u_{0}^{D'}
= u_{0}^{D}$ for any $D,\,D'$. The next equation in the recurrence
gives for $D,D' \geqslant 5$
\begin{equation} \label{red2a}
u_{1}^{D'} = u_{1}^{D}\frac{d-2}{d'-2}.
\end{equation}
Similarly for $D, \geqslant 7, \;D'=D+2$ one obtains:
\begin{equation} \label{red2b1}
u_{2}^{D+2} =u_{2}^{D}\frac{d-3}{d-1}.\end{equation} Continuing this
process further one  finds \be \label{red5}
-\frac{1}{d-1}\frac{\partial G_H^{D}}{\partial \sigma}= \ (2\pi)
G_H^{{\rm dir }\,D+2}, \ee where the ``direct'' part of the Hadamard
function is
\begin{equation}
G_{H\,{\rm dir}}=\frac{1}{(2\pi)^{d}}
\sum_{k=0}^{d-2}\frac{u_k}{\sigma^{d-1-k}}.
\end{equation}

\subsection{Divergencies}\label{4dim}
For the retarded Green's function one finds:\be  \label{red5a}
 G_{\rm ret} =\frac{1}{2(2\pi)^{d-1}}\Theta(x',\Sigma(x))
 \left( \sum_{m=0}^{d-2}\frac{(-1)^{m} u_{d-2-m}
\delta^{(m)}(\sigma)}{m!}- v \theta(- \sigma)\right). \ee
 The first term constitutes the direct part of the retarded function with
the support on the  light cone:
\begin{equation} \label{red5l}
 G_{\rm dir}  = \frac{1}{2(2\pi)^{d-1}}\Theta[\Sigma]
\sum_{m=0}^{d-2}\!\frac{(-1)^{m} u_{d-2-m}
\delta^{(m)}(\sigma)}{m!}.
\end{equation}

Similarly to the recurrent relations of the previous section, we
obtain for the retarded Green's functions:
\begin{equation} \label{red6}
\frac{\partial G_{\rm ret}^{D}}{\partial \sigma} =
-2\pi\left(d-1\right)G_{\rm dir}^{D+2}.
\end{equation}
Using the regularization  $\delta( \sigma)=\lim_{\varepsilon \to
+0}\delta( \sigma -\mathcal{E}),$ where
$\mathcal{E}=\varepsilon^2/2$, we obtain for the direct part of the
reaction force
\begin{equation}\label{red7a}
f^{\mu \; {\rm dir}}_{D+2}=-\frac{1}{D-1}\frac{\partial f^{\mu \;
{\rm dir}}_{D}(s,\mathcal{E})}{\partial \mathcal{E}}.
\end{equation}
The limit $\mathcal{E} \to +0$ has to be taken after the
differentiation. The direct force is due to the light cone part of
the retarded Green's function. This is not the full local
contribution to the Lorentz-Dirac force. An additional contribution
comes from the differentiation of the theta function in the tail
term $v  \, \theta( \sigma)$. In the scalar case this contribution
vanishes, but in the electromagnetic case an extra local term
arises:
\begin{equation} \label{rew2}
f^\mu_{\rm loc}=e^2\left([v_{\mu\alpha}]\dot{z}_{\nu}
-[v_{\nu\alpha}]\dot{z}_{\mu}\right)\dot{z}^{\nu}\dot{z}^{\alpha},
\end{equation}
where the coincidence limit  $[v_{\nu\alpha}]$ depends on the
dimension. The remaining contribution from the tail term will have
the form of an integral along the past half of the particle world
line.

The direct part of the Lorentz-Dirac force contains divergences. To
separate the divergent terms one can use the decomposition of the
retarded potential suggested in the case of four dimensions by
Detweiler and Whiting \cite{Detw02,Detw05}. In higher even
dimensions we can follow essentially the same procedure. We define
the ``singular'' part $G_{\rm S} $ of the retarded Green's function
as the sum of the symmetric part (self) and the tail function $v$ as
follows \be  \label{det0} G_{\rm S}(x,x')  = G_{\rm
self}(x,x')+\frac{v(x,x')}{4(2\pi)^{d-1}}= G_{\rm
self\;dir}(x,x')+\frac{v(x,x')\, \theta(-\sigma)}{4(2\pi)^{d-1}} .
\ee
 Here the direct part of the self function means its part
without the tail $v$-term. The remaining part of the Green's
function $ G_{\rm R}(x,x')=G_{\rm ret}(x,x')-G_{\rm S}(x,x')
$
 satisfies a
free wave equation and is regular.
 Taking into
account $\Box v=0$, it is clear that $G_{\rm S}$ satisfies the same
inhomogeneous equation as $G_{\rm self}$. The $v$-term in the second
line of Eq. (\ref{det0}) is localized outside the light cone.
Therefore the corresponding field (for instance, scalar), at an
arbitrary point $x$ will be given by
\begin{equation} \label{det1}
\phi_{\rm S} (x) =\phi_{\rm self\;dir}+\frac{m_0 q  }{4(2\pi)^{d-1}}
\int \limits_{\tau_{\rm ret}}^{\tau_{\rm adv}} v(x,z(\tau))d\tau,
\end{equation}
where the retarded and advanced proper time values $\tau_{\rm
ret}(x),\, \tau_{\rm adv}(x)$ are the intersection points of the
past and future light cones centered at $x$ with the world line.

Inserting (\ref{det1}) into the Lorentz-Dirac force defined on the
world-line $x=z(\tau)$, we observe that the integral contribution
from the tail term   vanishes and so the divergent part is  entirely
given by  $\phi_{\rm self\;dir}.$ Using for  delta-functions  the
point-splitting regularization we find that all divergent terms
arise as negative powers of  $\varepsilon$. To prove the
existence of the counterterms we consider the interaction term in
the action substituting the field as the S-part of the retarded
solution to the wave equation. In the scalar case we will have:
\begin{equation}\!\!\!\!\!\!S_{\rm S}\!=\!\frac{1}2 \int G_{\rm
S}(x,x')\rho(x)\rho(x')\sqrt{g(z)g(z')}\,dx \,dx',
\end{equation}
where a factor one half is introduced to avoid double counting when
self-interaction is considered. Substituting the currents we get
 the  integral over the world-line.
Since the Green's function is localized on the light cone  we
expand the integrand in terms of the difference $t=\tau-\tau'$
around the point $z(\tau)$:
\begin{equation} \label{det3}
S_{\rm S}\sim \int d\tau \int \sum_{k,l}
B_{kl}(\tau)\delta^{(k)}(t^2-\varepsilon^2)t^l dt.
\end{equation}
Here  the coefficients $B_{kl}(z)$ depend on the curvature, while
the delta-functions are flat: $\delta^{(k)}(t^2-\varepsilon^2)$. By
virtue of parity, the  integrals with odd $l$ vanish, so only the
odd inverse powers of $\varepsilon$ will be present in the
expansion. Moreover, once we know the divergent terms in some
dimension $D$, we can obtain by differentiation all  divergent terms
in $D+2$, except for $1/\varepsilon$ term.  The linearly divergent
term corresponds to $l=2k$. The integral is equal to
\begin{equation*}
\int\limits_0^{\infty} \delta^{(k)}(t^2-\varepsilon^2)t^{2k}
dt=\frac{(-1)^{k}(2k-1)!!}{2^{k+1}}\frac{1}{\varepsilon}.
\end{equation*}
In four dimensions this term is unique. Applying our recurrence
chain  we obtain the inverse cubic divergence in six dimensions and
calculate again the linearly divergent term. Thus in $D=2d$
dimensions we will get $d-1$ divergent terms from which $d-2$ can be
obtained by the differentiation of the previous-dimensional
divergence, and the linearly divergent will be new. This linearly
divergent self-action term in the action will have generically the
form
\begin{equation*}
S_{\rm S}^{(-1)}\sim \frac{1}{\varepsilon}\int \sum_{k=0}^{k=?}
\frac{(-1)^{k}(2k-1)!!}{2^{k+1}} B_{k,2k}(\tau)  d\tau.
\end{equation*}
Here the coefficient functions are obtained taking the coincidence
limits of the two-point tensors involved in the expansion of the
Hadamard solution. They actually depend on the derivatives of the
world line embedding function $z(\tau)$ as well as the curvature
terms taken on the world line:
\begin{equation*}
B_{k,2k}(\tau)=B_{k,2k}(\dot{z},\ddot{z} ,..., R(z(\tau)),
R_{\mu\nu}(z(\tau)),...).
\end{equation*}

The vector case is technically the same, now one has to  expand in
in powers $t$ in the integrand of
$$S^{\rm S}=\frac12\int G^{\rm
S}_{\mu\alpha}(x,x')j^{\mu}(x)j^{\alpha}(x')\sqrt{g(z)g(z')}\,dx
\,dx'.$$

From this analysis it follows that in any dimension the highest
divergent term can be absorbed by the renormalization of the mass as
in the generating four dimensional case. To absorb the remaining
$d-2$ divergences one has to add to the initial action the sum of
$d-2$ counterterms depending on higher derivatives of the particle
velocity. Typically the counterterms depend on the Riemann tensor of
the background.

\subsection{Four dimensions}\label{4dim}

In four dimensions the scalar retarded Green's function contains a
single direct term localized on the light cone  and a tail term:
\begin{equation}\label{4ret}
 G_{\rm ret} = \frac{1}{4\pi} \theta [\Sigma (x),x']
 \left[\Delta ^{1/2} \delta (\sigma )-v \theta ( \sigma )\right].
\end{equation}
The retarded solution for the scalar field reads
\begin{equation} \label{scagree}
 \!\! \! \! \! \!  \phi_{\rm
ret}(x)=m_0 q \int\limits_{-\infty}^{\tau_{\rm ret}(x)}\left[
-\Delta ^{1/2} \delta (\sigma)+v \theta ( \sigma )\right] d\tau'.
\end{equation}
Differentiating this expression we obtain
%Substituting the retarded potential (\ref{scagree}) into the right
%hand side of the equation (\ref{sc}), we obtain for
$\phi_{\nu}=\partial_\nu \phi $ on the world line:
 \be  \phi_{\nu}(z(\tau)) =   m_0 q \int\limits_{-\infty}^{\tau}
 [-\Delta^{1/2}\delta' (\sigma ) \sigma_{\nu }  -\Delta
^{1/2}_{;\nu}\delta (\sigma )-v \delta(\sigma )\sigma _{\nu
}+v_{\nu}]\, d\tau ', \ee
 where integration is performed along the
past history of the particle. All the two-point functions are taken
on the world-line at the points
 $x=z(\tau)$ (observation  point) and
 $z'=z(\tau')$ (emission  point).

To compute  local contributions  from the terms proportional to
delta-functions and its derivative, it is enough to expand the
integrand in terms of  the difference $s=\Delta \tau=\tau -\tau '$
around the point $z(\tau)$.  The Taylor (covariant) expansion of the
fundamental biscalar $\sigma (z(\tau),z(\tau' ))$ is given by
\cite{christ76}:
\begin{equation}
 \sigma (z(\tau),z(\tau' ))= \sum_{k=0}^{\infty} \frac{1}{k!}D^k\sigma (\tau,\tau)
 (\tau-\tau')^k,
\end{equation}
where we denote as $Q(\tau,\tau')$ the quantity $Q(z(\tau)
,z(\tau'))$ for any  $Q$ taken on the world-line, and $D$ is a
covariant derivative along the world-line (also denoted by dot):
\begin{equation*}
D\sigma\equiv\dot \sigma = \sigma _{\alpha} \dot z^\alpha,
\;\;\;\;D^2\sigma\equiv\ddot \sigma = \sigma _{\alpha\beta } \dot
z^\alpha \dot z^\beta +
  \sigma _{\alpha}\ddot z^\alpha,
\end{equation*}
etc. Such an expansion exists since the difference $s=\tau -\tau'$
is a two-point scalar itself: this is the   integral from the scalar
function $\int (- \dot z^{2})^{1/2} ds,$ along the world-line from
$z(\tau)$ to $z(\tau')$. Taking the limits and using   $\dot
z^2(\tau)=-1$, we find:
\begin{equation}
\sigma (s)= - \frac{s ^2}{2}- \ddot z^2(\tau) \frac{s
^4}{24}+\mathcal{O}(s^5).
\end{equation}
To obtain an expansion of the derivative of $\sigma$ over $z^\mu
(\tau)$ one can expand $\sigma^ {\mu}(\tau, \tau -s)$ in powers of
$s$. This quantity transforms as a vector at  $z(\tau)$ and a scalar
at $ z(\tau ')$, this
\begin{equation}\label{signu}
\sigma ^{\mu}(s)=  s \left(\dot z^\mu -\ddot z^{\mu}\frac{s}{2}
+\dddot z^\mu\frac{s^2}{6} \right)+\mathcal{O}(s^4),
\end{equation}
where the index $\mu$ corresponds to the point $ z(\tau ) $: $\sigma
^{\mu}=\partial \sigma (z,z')/\partial z_{\mu}$. Recall that the
initial
 Greek indices  correspond to $z(\tau')$. The expansion of
$\delta(-\sigma)$ will read:
\begin{equation}
\delta(-\sigma)=\delta(s^2/2)+s^4\frac{\ddot z^2(\tau)}{24}
\delta'(s^2/2)+...
\end{equation}
 where the derivative of the delta-function is taken with respect to
the full argument. Since the most singular term is $\Delta
^{1/2}\delta' (\sigma ) \sigma_{\nu }$, the maximal order giving the
non-zero result after the integration is $s^3$. (Note, that in order
to use our dimensional recurrent relations to obtain a reaction
force in higher dimensions we should perform an expansion up to
higher orders in $s$.) Thus, with the required accuracy,
$\delta(-\sigma)=2\delta(s ^2),$ and all the integrals for the
delta-derivatives are the same as in the flat space-time. This
allows us to use the same regularization of the  delta-functions
with double roots $\delta(s ^2)$ by the point-splitting. Expanding
the biscalar $\Delta^{1/2}$ and its gradient at  $z$ we have:
\begin{equation}\label{utochka}
 \Delta^{1/2}= 1+\frac{s^2}{12}R_{\sigma  \tau }\dot z^\tau \dot
z^\sigma, \;\; \partial_\nu\Delta^{1/2}= \frac{s}{6} R_{\nu \tau
}\dot z^\tau.
\end{equation}
Combining all the contributions we obtain finally for the field
strength on the world-line: \be  \label{scalderiv}
  \phi_{\nu}\Big|_{z(\tau)} =  m_0q \left( \frac {1}{2 \varepsilon }\ddot z_{\nu}
-\frac {1}{3}\dddot z_{\nu}-\frac {1}{6} R_{\nu \tau }\dot z^\tau -
\frac{1}{6} R_{\gamma \delta }\dot z^\gamma \dot z^\delta \dot
z_\nu\ +\frac{1}{12}R \dot z_{\nu} + \int \limits_{\infty}^{\tau}
v_{\nu}\, d\tau '\right).\ee  After the renormalization of mass
 one recovers the familiar equation.

Similarly, in the electromagnetic case one fins the retarded Maxwell
tensor on the world line  $x=z(\tau)$: \be  \label{emintens}
 F^{\rm ret}_{\mu\nu}\Big|_{z(\tau)}=e \int \limits_{-\infty}^{\tau} [u_{\mu
\alpha; \,\nu} \delta (\sigma )+u_{\mu \alpha} \sigma_{\nu}\delta
'(\sigma )   + v_{\nu \alpha ; \, \mu}+ v_{\mu
\alpha}\sigma_{\nu}\delta (\sigma )-\{ \mu\leftrightarrow \nu
\}]\dot{z}^{\alpha}d\tau'. \ee Performing expansions on the
world-line, one easily recover the DeWitt-Brehme-Hobbs result \be
\label{emfin} f^{\mu}_{\rm em}=
e^2\left[-\frac{\ddot{z}^{\mu}}{2\varepsilon}+\Pi^{\mu\nu}
\left(\frac{2}{3}\dddot z_{\nu}+\frac{1}{3}R_{\nu\alpha}\dot
z^\alpha\right)  +\dot z^{\nu }(\tau )\int \limits_{-\infty }^{\tau
}(v^{\mu }_{\,\,\,\alpha;\nu} - v_{\nu \alpha }^{\quad;\mu })\dot
z^{\alpha } (\tau' )d\tau'\right], \ee where
$\Pi^{\mu\nu}=g^{\mu\nu}-{\dot z}^\mu {\dot z}^\nu$. For geodesic
motion the entire reaction force is given by the tail term.
\subsection{Self and rad forces in curved space-time}
Apart from the tail term, another new feature of the instantaneous
momentum balance between the radiating charge and radiation is the
presence of the finite contribution in the self part of the reaction
force originating from the half-sum of the retarded and advanced
potentials. As we have seen in the previous section, in the case of
Minkowski space the self part in four and other even dimensions is a
pure divergence which has to be absorbed by renormalization of the
mass and (in higher dimensions) the bare coupling constants in the
higher derivative counterterms. In curved space, as was first shown
in \cite{Dewi}, the tail term in the equation for radiating charge
moving along the geodesic with non-relativistic velocity (in weak
gravitational field)  contains apart from the dissipative term also
the conservative force. This conservative force was later found for
a static charge in the Schwarzschild metric \cite{SmWi}. Here we
would like to explore the significance of this result in view of the
above analysis of self/rad decomposition. In the paper \cite{DWDW}
the splitting of the retarded field into the self and rad parts was
not used. Considering the geodesic motion we have the equation:\be m
\ddot z^\alpha = e^2 \, \dot z^\beta \int_{-\infty }^{\tau}
v^{\alpha}_{{\rm ret}\beta \gamma } \dot z^{\gamma }(s')ds '.  \ee
  Splitting the tail function
with respect to T-parity \be v^{\alpha}_{{\rm ret}\beta \gamma }
\dot z^{\gamma }= v^{\alpha}_{{\rm self}\beta \gamma } \dot
z^{\gamma }+ v^{\alpha}_{{\rm rad}\beta \gamma } \dot z^{\gamma },
\ee and repeating the calculations along the lines of \cite{DWDW}
one finds in the weak-field non-relativistic case   the
corresponding (spatial) parts of the reaction force in terms of the
flat space theory:\be{\bf f}_{\rm self}={\bf f}_{\rm div}+{\bf
f}_{\rm WWSM},\qquad  {\bf f}_{\rm rad}= {\bf f}_{\rm Schott },\ee
where the divergent term is the same as in the flat space. The two
finite terms  \be{\bf f}_{\rm WWSM}=\frac{GMe^2}{r^4}{\bf r},\qquad
\qquad{\bf f}_{\rm Schott}=\frac23 e^2{\bf {\dot a}} \ee are the
 WWSW  force and the Schott force. Therefore, the WWSW force is a
finite part of the T-even (self) contribution to the  tail, while in
the T-odd (rad) part reproduces the  Schott term which is precisely
the same  as in the flat space. Similar result holds for the scalar
radiation.

\section{Gravitational radiation}

Gravitational radiation in the framework of linearized gravity on
the curved background may seem similar to scalar or electromagnetic
radiation, but the similarity is incomplete. The difference is that
radiation of non-gravitational nature emitted by bodies moving along
geodesics in the fixed background  is  not influenced by the
non-linear nature of the Einstein equations.  For gravitational
radiation this is not so.  The scalar or vector linear field
equations in curved space imply vanishing of the covariant
divergence of the field stress tensor of matter field with respect
to the background metric. In the gravitational case one has the
Bianchi identity which generally does not imply the covariant
conservation of the matter perturbation stress-tensor alone unless
the background is vacuum. In most of the literature on gravitational
radiation reaction the background is assumed to be vacuum. This
seems to be satisfactory for the Schwarzschild or Kerr metrics. But
physically we have to deal not with an eternal black hole, but with
the collapsing body which is not globally vacuous. Meanwhile the
conservation laws which may hold in the asymptotically flat case has
to be considered globally. It turns our that we must take into
account contribution from the (perturbed) source of the background
field as well. This is directly implied by the Bianchi identities.
\subsection{Bianchi identity}
 Let the
background be generated by the stress-tensor
 $\stackrel{B}{T}{\!\!}^{\mu\nu}$. We are interested by gravitational
radiation emitted by a point particle of mass m moving along the
geodesic of the background. Perturbations caused by the particle are
assumed to be small so the particle stress-tensor \be
\label{partic}\stackrel{m}{T}{\!\!}^{\mu\nu}= m \int \dot z^{\mu
}(\tau ) \dot z^{\nu }(\tau )\frac{\delta(x-z(\tau))}{\sqrt{-g}}
d\tau \ee is the first order quantity with respect to
$\stackrel{B}{T}{\!\!}^{\mu\nu}$. By construction, the tensor
(\ref{partic}) is divergence-free provided the particle follows the
geodesic in the space-time. Since it is the first order quantity, it
must be divergence-free with respect to the background covariant
derivative up to the terms of the second order. Expanding the full
metric $\hat{g}_{\mu \nu }=g_{\mu \nu } +\kappa h_{\mu \nu }$, where
$\kappa^2=8\pi G$, and the Einstein tensor \be {\hat
G}^{\mu\nu}=\stackrel{B}{G}{\!\!}^{\mu\nu}
+\stackrel{1}{G}{\!\!}^{\mu\nu},\ee we could naively expect the full
Einstein equations to be of the form \be {\hat
G}^{\mu\nu}=\kappa^2\left( \stackrel{B}{T}{\!\!}^{\mu\nu}
+\stackrel{1}{T}{\!\!}^{\mu\nu}\right).\vspace{-.3cm} \ee Since $
\stackrel{B}{G}{\!\!}^{\mu\nu}=
\kappa^2\stackrel{B}{T}{\!\!}^{\mu\nu}, $ we then should have \be
\label{equi} \stackrel{1}{G}{\!\!}^{\mu\nu}=
\kappa^2\stackrel{m}{T}{\!\!}^{\mu\nu}. \ee  But the left hand side
of this equation is not divergence-free with respect to the
background covariant  derivative. Expanding the full covariant
derivative as \vspace{-.2cm}\be \hat{\nabla}{\!\!}_\mu=
\stackrel{B}{\nabla}{\!\!}_\mu +\stackrel{1}{ \nabla}{\!\!}_\mu
\vspace{-.3cm}\ee and taking into account the Bianchi identity for
the background
$\stackrel{B}{\nabla}{\!\!}_\mu\stackrel{B}{G}{\!\!}^{\mu\nu}=0$ we
obtain in the first order \be
\stackrel{B}{\nabla}{\!\!}_\mu\stackrel{1}{G}{\!\!}^{\mu\nu}=
-\stackrel{1}{\nabla}{\!\!}_\mu\stackrel{B}{G}{\!\!}^{\mu\nu}.\ee
Therefore \be
\stackrel{B}{\nabla}{\!\!}_\mu\stackrel{m}{T}{\!\!}^{\mu\nu}=
-\stackrel{1}{\nabla}{\!\!}_\mu\stackrel{B}{T}{\!\!}^{\mu\nu},\ee
where the right hand side is the first order quantity. Thus Eq.
(\ref{equi}) is contradictory. Physically the reason is that we have
to take into account the perturbation of the background $\delta
T^{\mu\nu}$ caused by the particle, so that the correct equation
should be \be \label{true} \stackrel{1}{G}{\!\!}^{\mu\nu}=
\kappa^2\left(\stackrel{m}{T}{\!\!}^{\mu\nu}+\delta T^{\mu\nu}
\right). \ee  But this is not an equation for $h_{\mu\nu}$, since in
order to find $\delta T^{\mu\nu}$ one has to consider the matter
field equations for the background metric. The problem thus becomes
essentially non-local. This non-locality does not reduce to that of
the tail term in the DeWitt-Brehme equation.

\subsection{Vacuum background}\vspace{-.3cm}
If $\stackrel{B}{T}{\!\!}^{\mu\nu}=0$ the above obstacle is removed
and one can proceed further with the linearized equations for
$h_{\mu\nu}$. The derivation of the reaction force  initiated in
\cite{morette} and completed in \cite{mino} was based on the
DeWitt-Brehme type calculation involving the integration of the
field momentum over the small tube surrounding the particle
world-line. As was noted later  \cite{SaPo}, this derivation had
some drawbacks (for more recent discussion and further references
see \cite{GraWa}. One problem consisted in computing the
contributions of "caps" at the ends of the chosen tube segment which
were not rigorously calculated. Another problem was the singular
integral over the internal boundary of the tube which was simply
discarded. In addition, the usual mass-renormalization is not
directly applicable in the gravitational case: due to the
equivalence principle the mass does not enter into the geodesic
equations.

In \cite{gaspist} the local derivation of the gravitational reaction
force was given which is free from the above problems and, in
addition, is much simpler technically. It deal with the quantities
defined only on the world line and does not involve the ambiguous
volume integrals over the world-tube at all.  The elimination of
divergences amounts to the redefinition of the affine parameter on
the world-line.

 We start with reparametrization invariant form of the particle action
introducing the einbein  $e(\tau )$ on the world line acting as a
Lagrange multiplier:
\begin{align}\label {act}
S[ z^\mu, e]=-\frac{1}{2}\int \left[ e (\tau )\, g_{\mu \nu } \dot
z^{\mu } \dot z^{\nu } +\frac{m^2}{e(\tau )}\right]d\tau.
\end{align}
Variation with respect to $z^\mu (\lambda )$ and $e(\tau )$ gives
the equations
\begin{align}\label{mot}
\frac{D}{d\tau }(e\dot z^\mu )=0,\quad e=\frac{m}{\sqrt{ -g_{\mu \nu
} \dot z^\mu \dot z^\nu } },
\end{align}
and we obtain the geodesic equation in a manifestly
reparametrization invariant form
\begin{align}\label{caf}
\frac{D}{d\tau }\left(\frac{\dot z^\lambda}{\sqrt{ - g_{\mu \nu }
\dot z^\mu \dot z^\nu }}\right)=0.
\end{align}
  Assuming now that the particle motion with no
account for radiation reaction is geodesic on the background metric,
the perturbed equation in the leading order in $\kappa$ will read
\begin{align}\label{prop}
\ddot z^\mu = \frac{\kappa }{2}\left(g^{\mu \nu }-\frac{\dot z^\mu
\dot z^\nu}{\dot z^2}\right) ( h_{\lambda \rho ;\nu}-2h_{\nu \lambda
;\rho}) \dot z^\lambda \dot z^\rho,
\end{align}
where  contractions are with the background metric.

The particle energy momentum tensor in our formulation will read \be
 \stackrel{m}{T}{\!\!}^{\mu\nu}=   \int e(\tau) \dot z^{\mu }(\tau ) \dot
z^{\nu }(\tau )\frac{\delta(x-z(\tau))}{\sqrt{-g}} d\tau, \ee and we
choose the non-perturbed ein-bein $e_0=$const as a bare parameter.
After the calculation similar to that of the preceding section we
obtain \bea \ddot z^\mu &=& \kappa^2 e_0 \Bigl\{ \frac{7}{2\epsilon
} \ddot z^\mu + \frac{1}{4}\Pi^{\mu\nu}\int \limits_{-\infty }^\tau
\Bigl[4 v _{\nu \lambda\alpha \beta ;\rho}- 2( g_{\nu \lambda }
v^\sigma_{\sigma \alpha \beta ;\rho}+v_{\lambda\rho \alpha \beta
;\nu})- \nonumber\\ &-& g_{\lambda \rho } v^\sigma_{\sigma \alpha
\beta ;\nu})\Bigr] \dot z'^{\alpha }\dot z'^{\beta } \dot z^\lambda
\dot z^\rho d\tau' \Bigr\}. \eea

Renormalization of the einbein is
\begin{equation} \label{renorm}
\biggl( \frac{1}{e_0}-\frac{7\kappa ^2}{2\epsilon } \biggr) \ddot
z^\mu   = \frac {1}{e}\ddot z^\mu.
\end{equation}
Finally we choose the renormalized affine parameter so that $\dot
z^2=-1$, which is equivalent to set  $e=m$ and obtain the MiSaTaQuWa
equation.  As was shown in \cite{gaspist} this equation remains
valid in a class on non-vacuum metrics, in particular, for Einstein
spaces. \subsection{Gravitational radiation for non-geodesic motion}
If the particle world-line is non-geodesic, the radiation reaction
force contains a putative antidamping term which is  a local part of
the  {\em rad}  contribution to the
 self-force: \be  f^\mu_{\rm{ rad}}  = -\frac{11\kappa^2}{3}
  (g^{\mu \nu }-{\dot z^\mu \dot z^\nu} )
 \dddot z_\nu\;\;\;+\;\;\;{\rm tail}.\ee
The reason  is simply that the source of gravitational radiation
 is incomplete and the stress tensor is not divergence-free as required.
Indeed, if  the force is non-gravitational, one has to  take
 into account the contribution of stresses of the field causing the
 body to accelerate. For instance, to describe gravitational
 radiation of an electron in the atom, one has to add the contribution form
  the Maxwell field
 stresses (spatial components, non-relativistic motion):\be\Box \psi^{ij} =
 -\kappa^2 G T^{ij},\qquad   T^{ij}=\stackrel{m}{T}{\!\!}^{ij}+
 \stackrel{st}{T}{\!\!}^{ij},\ee
 where  \be \stackrel{m}{T}{\!\!}^{ij}=\sum_{a=1,2}m_a {\dot z}^i_a {\dot
 z}^j_a\delta^3(X_a),\quad
 \stackrel{st}{T}{\!\!}^{ij}=
 -\frac{e_1 e_2}{4\pi}\frac{X^i_1X^j_2}{(X_1^2 X_2^2)^{3/2}} + (i\leftrightarrow j),\ee
 and $X^i_a=x^i-z^i_a(t).$ Using
 this source one can calculate the gravitational force   and find
 the gravitational Schott term \be
 { f^i_{\rm Gshott}}=-\frac{G\mu}{15}\frac{d^5D^{ij}}{dt^5}x_j,\quad
\mu=\frac{m_1m_2}{m_1+m_2},\ee where $D^{ij}$ is the quadrupole
moment.

Note that this  derivation of the Schott term is not based on the
local calculation, the two-body treatment was necessary. This
features seems to be general: gravitational radiation reaction form
non-geodesically moving particle can not be described by some
DeWitt-Brehme-type equation.

 \section{Conclusions}
The purpose of this lecture was to discuss some subtle points
associated with interpretation of the radiation reaction force. We
have shown that the Lorentz-Dirac equation in classical
electrodynamics describes the balance of three and not just two
momenta: the mechanical momentum of the particle, the momentum of
emitted radiation,  and the momentum carried by the electromagnetic
field bound to the charge. The total momentum is conserved, but this
does not imply an instantaneous balance of the emitted momentum and
that of the particle. The bound field momentum described by the
Schott terms destroys the local balance. The total balance, however,
is restored if one consider the situation when the charge has zero
acceleration at the initial and final  moments, or for a periodic
motion subject to averaging. These considerations are equally
applicable to radiation reaction of a charge in curved space-time
and for gravitational radiation reaction. This explains, in
particular, the necessity of averaging in calculating the evolution
of the Carter constant in the Kerr field \cite{Ta05}.

A novel feature related to curved space is the existence of the
finite WWSW force arising due to the tidal deformation of the bound
electromagnetic field of the charge. This force is often interpreted
as part of radiation reaction force, but one has clearly understand,
however, that it has nothing to do  either with radiation  or with
the Schott force. As we have shown, it is given by the T-even part
of the retarded field, and thus present a finite remnant from the
mass renormalization.

Derivation of the reaction force of non-gravitational nature acting
on a charge moving (both geodesically and non-geodesically) in
curved space-time can be computed directly substituting the retarded
field into the equations of motion, as is the Minkowski space. The
regularization is easily achieved by the point-spliting, and
divergences are eliminated by renormalization of mass. In higher
dimensions one needs counterterms depending on higher derivatives of
the velocity. Divergencies may contain the Riemann tensor of the
background.

Gravitational radiation reaction force can be obtained in a way
similar to non-gravitational one only in vacuum space-time. In
non-vacuum background the source of radiation apart from the local
contribution from the particle must contain the contribution from
the perturbed background. This can be seen from the analysis of the
Bicnchi identity. This second contribution is non-local, so the
possibility to obtain the equation of the DeWitt-Brehme type seems
implausible.

For a non-geodesically moving mass the formal derivation of the
reaction force leads to putative antidamping effect. To cure this
problem one has to take into account the contribution of stresses
forcing the mass to accelerate. Then in the non-relativistic case
one derives the gravitational quadrupole Schott term, but the
derivation is non-local. This is another example when the
(quasi)local equation of motion with the reaction force does not
exist. Here by quasilocality we mean the possibility of the tail
term.

\section*{Acknowledgments}
The author is grateful to the Organizing Committee for  invitation
and support in Orlean. Useful discussions with the participants of
the School on Mass and Capra conference are acknowledged. Most of
the results presented in this lecture were obtained in collaboration
with Pavel Spirin, to whom the author in indebted. The work was
supported by the RFBR project 08-02-01398-a.

\end{document}